\documentclass[]{llncs}
\usepackage{setspace}
\usepackage{multirow}
\usepackage{hyperref}
\usepackage{alltt}
\newcommand{\rimp}{\Rightarrow}
\newcommand{\status} {{\tt status }}

\newcommand{\abort}{{\it abort}}
\newcommand{\AG}{{\textbf{ AG}}}
\newcommand{\AF}{{\textbf{ AF}}}

\newcommand{\no}{{\it no}}
\newcommand{\yes}{{\it yes}}
\newcommand{\commit}{{\it commit}}

\newcommand{\R}{{\cal R}}
\newcommand{\U}{{\cal U}}
\newcommand{\E}{{\cal E}}
\newcommand{\Prop}{{\it Prop}}
\usepackage{amsfonts} 
\usepackage{graphicx}
\usepackage[]{amsmath}

\title{Verifying Real-time Commit Protocols Using Dense-time Model Checking Technology
\thanks {The first and second authors were supported by ARC discovery project grant DP110104669. } }
\author{Omar I. Al Bataineh,
 Mark Reynolds, 
 Tim French, 
 and Terry Woodings}
\institute {University of Western Australia} 
\date{}
\begin{document}

\maketitle

\begin{abstract}
The timed-based automata model, introduced by Alur and Dill, provides
a useful formalism for describing real-time systems. Over the last two decades,
several dense-time model checking tools
have been developed based on that model. The paper 
considers the verification of real-time distributed commit protocols 
using dense-time model checking technology.
More precisely, we model and verify the well-known timed two phase commit protocol
in three different state-of-the-art real-time model checkers: UPPAAL, Rabbit, and RED,
and compare the results.

\end{abstract}

\section{Introduction}

Real time systems are currently being used in many safety critical applications such as computer-controlled
medical devices, air traffic control systems, and real-time database systems.
Ensuring the correctness of real-time systems is a challenging 
task. This is mainly because the correctness of real-time systems 
depends on the actual times at which events occur. 
 Moreover, these systems involve interactions
of a number of concurrent components that have a high level of complexity.
Such interactions might lead to many subtle or undesirable situations if they are not considered carefully.
Hence, real-time systems need to be rigorously modeled and verified in order to
formally prove their correctness with respect to the desired properties. 
It is well-known that only formal methods of verifications can guarantee that
the system\ meets its desired properties.
One of the well-known formal verification techniques is the model checking technique \cite{CGP99}.

Model checking is an automated
method for verifying finite state systems. 
Model checking has been successfully used to find non-trivial errors in hardware designs, 
distributed systems, and security protocols.
Applying model checking to prove the correctness of a system design consists of
three formal independent steps, which are modeling, specification, and verification.
Model checking uses a variety of sophisticated heuristics and symbolic implementation techniques
to check that a logical formula holds in the given system design.
Typically, formulas are expressed using logics such as temporal logic, belief logic, or epistemic
logic, while the system will be expressed as a set of states (finite number of states)
and a set of transitions. When the system fails to meet a desired property, the model
checker produces a counterexample that helps to identify the source of the error in
the system design.

 Because of time constraints in real-time systems, traditional model checking approaches
 based on finite state automata and temporal logic are not sufficient.
 Since they can not capture the time requirements of real-time systems upon
 which the correctness of these systems relies.
 Developing formal methods for verifying real-time systems
 has been an active area of research. 
 Several researchers have proposed different modeling formalisms
 for describing real-time systems such as
 timed transition systems \cite{Larsen97}, timed I/O automata \cite{kaynar03}, 
 and timed automata model \cite{Alur94}. 
 Although a number of formalisms have been proposed,
the \textit{timed automata model} of Alur,
Courcoubetis, and Dill \cite{Alur94} has become the standard.
In fact the majority of the existing dense-time model checking tools in the literature are based on this model.
Examples include UPPAAL \cite{Beh04}, KRONOS \cite{Daws96}, RED \cite{wang04}, and Rabbit \cite{Beyer03}. 

In this paper we consider the application of dense-timed model checking
to real-time distributed commit protocols.
In general, commit protocols can be classified
into conventional (non-real-time) commit protocols
and real-time commit protocols.
In conventional commit protocols such as the classical two phase commit protocol (2PC) \cite{Ber87},
processes need to coordinate with each other
in order to reach a common decision on whether or not to commit the 
transaction, but there is no strict or hard time constraints on agreement.
On the other hand, in real-time commit protocols, all processes should commit the
transaction before the deadline expires or all of them
should abort immediately upon deadline expiry.
So real-time transaction protocols should satisfy data consistency as well as
they must satisfy timing constraints associated with transaction.
Model checking has frequently been 
used to verify conventional commit protocols \cite{jeff,atif09,Jan92,Olve08}. 
In comparison, little work has been done on model checking 
real-time commit protocols. 

In this contribution, we conduct a comparative study of a number of model checking tools, based on
a variety of approaches to representing real-time systems.
We have selected one specific practical real-time commit protocol, the \textit{timed two phase commit protocol} \cite{Dav89},
implemented it in quite different `dense' timed model checkers, and verified its relevant 
properties. Specifically, we consider the model checkers UPPAAL, Rabbit and RED. 
UPPAAL deals with the logic of TCTL \cite{Alur93} using difference bound matrices (DBM) based algorithm.
Rabbit is a model checker based on timed automata extended with concepts for modular modeling
and performs reachability analysis using BDD based algorithm.
RED is a model checker with dense-time models based on 
CRD (Clock-Restriction Diagrams) algorithm.
Analysing the protocol using these tools requires us to learn the modelling and 
the specification language of the tools,
which allows us to discover the capabilities and the limitations of these tools.
Moreover, it allows us to compare the efficiency and the performance of the algorithms 
and data structures implemented in these tools.

All three implementations were carried out on the same verification server in 
order to obtain results that can be used to compare the performance of these tools. 
RED outperformed both UPPAAL and Rabbit in terms of performance, scalability,
and expressivity of its specification language.  
Unlike UPPAAL and Rabbit, RED supports full TCTL language and allows verifying 
formulas with nested temporal modalities.
Heuristics used in RED allowed us to verify the protocol for a large number 
of processes and great number of clocks.
During our analysis we have been able to discover
some interesting conclusions about the protocol. In particular,
we find that the central property of the protocol, the data consistency property,
might be temporarily broken during the execution of the protocol.
However, when the protocol does terminate, we find that the consistency is preserved.

The structure of the paper as follows. Section \ref{sec:two} describes
the concept of a real-time distributed system and gives
an informal description for the timed two phase commit protocol.
Section \ref{sec:three} describes briefly the theory of timed automata and
the timed temporal logic.
Sections \ref{sec:four} to \ref{sec:six} are dedicated to the implementation of the protocol
in the model checkers, respectively, UPPAAL, Rabbit, and RED.
We discuss correctness conditions of the protocol in Section \ref{sec:seven}.
In Section \ref{sec:eight} we formalise the specifications of the protocol
in the input language of each model checker. 
Section \ref{sec: results} discusses the verification results of the protocol.
Section \ref{sec:comparison} compares the performance results of the three tools. 
We discuss related work in Section \ref{sec:relatedwork} and conclude in Section \ref{sec:conclusion}.

\section{Real-time Distributed Transaction Protocols} 
\label{sec:two} 

Real-time database systems are transaction processing systems that are
designed to handle transactions that have real-time constraints (deadlines).
These systems can be said to be failed if they can not
complete within their deadline.
In distributed real-time transaction protocols, all processes should commit the
transaction before the deadline expires or all of them should abort immediately
upon deadline expiry. 
There are many database applications that have real-time constraints.
Examples include web-based auctions, medical records, 
reservation systems, and banking information systems.
Several commit protocols have been proposed to handle real-time
transactions such as Timed Two Phase Commit (T2PC) protocol \cite{Dav89},
PROMPT commit protocol \cite{prompt}, the deadline-driven conflict resolution 
protocol \cite{lam}, and the double space commit protocol \cite{Qin}.
This paper analyses the T2PC protocol as a case
study on real-time commit protocols using three state-of-the-art 
dense-timed model checkers. 

The T2PC protocol aims to maintain data consistency of all distributed
database systems
as well as having to satisfy the
time constraints of the transaction under processing.
The protocol is mainly based
on the well-known two phase commit (2PC) protocol, 
but it incorporates several intermediate
deadlines in order to be able to handle real-time transactions.
We describe first the basic 2PC protocol (without deadlines) and then discuss
how it can be modified to be used for real-time transactions.
The 2PC protocol can be summarised
as follows \cite{Ber87}.

A set of processes $\{p_{1},.., p_{n}\}$ prepare to involve in a distributed transaction. 
Each process has been given its own subtransaction.
One of the processes will act as a coordinator and all other
processes are participants.
The protocol proceeds into two phase.
In the first phase (voting phase),
the coordinator broadcasts a start message
to all the participants, and
then waits to receive vote messages from the participants.
The participant
will vote to commit the transaction if all its local computations
regarding the transaction have been completed successfully;
otherwise, it will vote to abort.
In the second phase (commit phase), if the coordinator
received the votes of all the participants, 
it decides and broadcasts the decision.
If all the votes are `yes' then the coordinator will
decide to commit the result of the transaction. However, if one
vote said `no', then the coordinator will decide to abort the transaction.
After sending the decision, the coordinator waits to receive a COMPLETION messages from all the participants.
The protocol terminates whenever the coordinator receives all COMPLETION messages successfully.

For the above 2PC protocol, let $S$ be the absolute start time of the protocol, 
and $D$ be the absolute deadline. For each of the participants,
let $t_{i}$ be the maximum time needed to receive a decision message 
from the coordinator, perform the commit or abort action,
and send a completion message to the coordinator. We will call the largest
of all the $t_{i}$ the $\tau_{max}$.
For the coordinator, let $\tau_{d}$ be the maximum amount of time
needed to receive the votes of all the participants, process them,
and make a decision; and $\tau_{f}$ be the maximum amount of time needed
to receive the completion signal from all the participants.
We will use $\Delta$ to denote the bound on execution of \textit{send},
and $ \Delta^{*}$ to denote the bound on execution of broadcast (\textit{send-all}).

The following intermediate deadlines have been added to the basic 2PC protocol
in order to be able to handle real-time transactions \cite{Dav89}:

\begin{itemize}
\item $D_{p} = D - \Delta - \tau_{f} $: deadline for sending a completion message by a participant.
Each participant must complete the commit or abort action and send the completion message within $\Delta$ time unit,
so that the coordinator still has enough time to process it with at most $\tau_{f}$ time units before the
expiry of $D$.
\item $DEC = D_{p} - \tau_{max} - \Delta^{*}$: deadline for sending a decision by the coordinator.
This represents the maximum message delay for the broadcast decision to arrive the participant.

\item $V = DEC - \Delta - \tau_{d}$: deadline for a participant to vote.
The participant must vote and send the vote to the coordinator before the $DEC$
timer expires.

\item $[LST_{i}, D_{p}]$: the time interval during which $P_{i}$ requests a reserve of $t_{i}$
 time units of resources needed to perform the decided-upon action.
 
\end{itemize}

Note that the correctness of the T2PC protocol depends mainly on the way we select the
values of the above timing parameters. In particular, the coordinator
should choose the value of $D$ to be sufficiently long to allow the
participants to receive the start message and return the completion message
in time for the coordinator to determine the result. 
The correctness of the protocol depends also on a condition that a fair scheduling policy is imposed,
this condition is necessary in order to avoid situations in which some participants
may miss the deadline if they schedule to execute until after the deadline $D$. 

The basic 2PC protocol (without deadlines and timers) has been extensively studied and 
analysed using model checking \cite{jeff,atif09,Jan92,Olve08}.   
However, no work has been done on 
model checking the real-time version of the protocol (T2PC).
Of course, analysing real-time commit
protocols is much harder than analysing
conventional commit protocols since real-time 
protocols usually involve many timers in their design which increase 
the algorithmic complexity of the analysis.

\section{Timed Automata and Real-time Temporal Logic}
\label{sec:three} 
In this section we describe briefly the semantics of timed automata
and then discuss the syntax and the semantics of TCTL logic. 

Timed automata are an extension
of the classical finite state automata with clock variables to model timing
aspects \cite{Alur94}. 
The theory of timed automata provides a powerful
formalism to model and verify real-time systems.
Systems whose correctness depend on strict timing constraints
such as response time, deadlines, communication delay 
can be easily described using the timed automata model.
Research on timed automata theory has been an intensive field of research
since it was introduced in 1990. 
Although a number of formalisms have been proposed to model timed systems,
the timed automata model has become the standard.
We therefore find that the majority of the existing real-time model 
checking tools in the literature are based on this model.
Examples include UPPAAL \cite{Beh04}, KRONOS \cite{Daws96}, RED \cite{wang04}, and Rabbit \cite{Beyer03}.

We begin by briefly reviewing some basic definitions of the timed automaton
model proposed by Alur and Dill \cite{Alur94} and the transition systems that underlies 
its semantics.

\begin{definition}
Given a set $X$ of clock variables, the set $\Phi(X)$ of clock constraints
$\phi$ is defined by the following grammar
\[
\phi ::= x \sim c  \mid \phi_{1} \land \phi_{2}
\]
where $x \in X$,
$c \in \mathbb{N}$, and $\sim \in \{<, \leq, =, >, \geq \}$.
\qed
\end{definition} 

A clock interpretation $v$ for a set $X$ is a mapping from
$X$ to $\mathbb{R}^{+}$ where $\mathbb{R}^{+}$ denotes the set 
of nonnegative real numbers including zero.  

\begin{definition}
A timed automaton $A$ is a tuple $(\Sigma, S, S_{0}, X, E)$, where

\begin{itemize}

\item $\Sigma$ is a finite set of actions.

\item $S$ is a finite set of states.

\item $S_{0}$ is a finite set of initial states.

\item $X$ is a finite set of clocks.

\item $E \subseteq S \times S \times \Sigma \times 2^{X} \times \Phi(C)$
is a finite set of transitions.
An edge $(s, s^{'}, a, \lambda, \sigma)$ represents
a transition from state $s$ to state $s^{'}$ after performing action $a$.
The set $\lambda \subseteq C$ gives the clocks to be reset with this transition,
and $\sigma$ is a clock constraint over $C$.   \qed
\end{itemize}
\end{definition}

The semantics of a timed automaton $(\Sigma, S, S_{0}, X, E)$ is defined by 
associating a transition systems with it. With each transition a clock constraint is associated. 
The transition can be taken only if the clock constraint on the transition is satisfied.
There are two basic types of transitions:
\begin{enumerate}
\item delay transitions that model the elapse of time while staying at some location,
\item action transitions that execute an edge of the automata.
\end{enumerate}
A state $ (l, v)$ or configuration consists
of the current location and the current values of clocks.
The initial state is $ (l_{0}, v_{0})$ where $ v_{0}(x) = 0$ for all $x \in X$.
A timed action is a pair $(t, a)$ where $a\in\Sigma$ is an action performed by an Automata $A$
after $t \in \mathbb{R}^{+}$ time units since $A$ has been started. 
Complex systems can be described as a set of timed automata executing in parallel.
A network of timed automata can be expressed as $ A =  \langle A_{1}, ..., A_{2}\rangle$
where $A_{i}$ is a timed automaton.

\begin{definition}
An execution of a timed automata
$A = (\Sigma, S, S_{0}, X, E)$ with an initial state $ (l_{0}, v_{0})$
over a timed trace $ \zeta= (t_{1}, a_{1}), (t_{2}, a_{2}),(t_{3}, a_{3}),..$ is a sequence of transitions:
$$\langle l_{0}, v_{0} \rangle \xrightarrow{d1} \xrightarrow{a1} \langle l_{1}, 
v_{1}\rangle \xrightarrow{d12} \xrightarrow{a2} \langle l_{2}, 
 v_{2} \rangle \xrightarrow{d3} \xrightarrow{a3} \langle l_{3}, v_{3} \rangle ...$$
satisfying the condition $t_{i} = t_{i-1}+d_{i}$ for all $i \geq 1$.
\qed
\end{definition}

The transition system in timed automata is infinite as clocks are real-valued,
therefore timed automata models can not be directly model checked. 
However, there exist methods to reduce the infinite state space of timed systems 
to finite space while preserving properties of interest.
Examples of abstraction methods include the region \cite{Alur90} and zone \cite{Alur98} methods. 

Analysing the behaviour of timed systems requires to verify
whether they satisfy the timing properties that they are supposed to satisfy. 
Verification techniques based on classical temporal logic are inadequate for real-time systems
since they lack the expressiveness to capture quantitative temporal constraints
upon which the correctness of these systems relies.
An easy and effective way to allow the verification of dense-time properties 
is to add bounds in the CTL temporal operators. The extended 
logic is called TCTL \cite{Alur93}. 
The expressive power of TCTL is similiar
to the CTL but it incorporates time-constrained modalities in order to specify
timing properties. Using this logic we can specify properties
 like event $e_{1}$ has to occur after 10 time units after
 event $e_{2}$ occurred.
The syntax of TCTL logic can be given by
the following definition.

\begin{definition} \textbf{(Syntax of timed computation tree logic)}
Let $A$ be a timed automaton with set of clocks $X$ and set of propositions
$\Prop$, and let $Z$ be a set of clocks disjoint with $X$.
The set of formulas of TCTL is defined by the following grammar:
\[
\psi ::= p \mid \phi \mid \neg \psi \mid \psi_{1} \lor \psi_{2} \mid \textbf{E} [\psi_{1} \textbf{U} \psi_{2}] \mid 
\textbf{A} [\psi_{1} \textbf{U} \psi_{2}]
\]
where $p \in \Prop$, $z \in Z$, and $\phi$ is a clock constraint as defined in Definition 1. \qed

\end{definition}

Before we discucss the formal semantics of TCTL, let us introduce some 
notations. 
A position $m$ of a run $r\in \R_{A}$ is a pair $(i, \sigma ) \in \mathbb{N} \times \mathbb{R^{+}}$. 
The set of positions of the form $(i,\sigma)$ characterizes the
set of states through which the run $r$ passes while time flows from state $q_{i}$
to state $q_{i+1}$. We use the notation $\sim$ to refer to one of the following
relations $\{<, \leq, =, >, \geq \}$.
The formulas of TCTL are interpreted over the states of a model.
The propositions and the clock variables of a TCTL formula $\varphi$
are evaluated in states.

\begin{definition} \textbf{(Semantics of TCTL)} 
Let $M$ be a model and let $q \in \Sigma_{M}$ be a state that is reachable
in $M$. The state $q$ satisfies the TCTL formula $\varphi$ in $M$, denoted by 
$q \models_{M} \varphi$, if $q \models_{M, \emptyset} \varphi$
for the empty clock environment $\emptyset$.
We define the semantics of the logic by means of a satisfaction relation $M \models \varphi$. 
This satisfaction relation is defined inductively as follows:

\begin{itemize}

\item $ q \models x \sim c$ ~~~~~~~~~ iff $v(x) \sim c$
\item  $ q \models x - y \sim c$ ~~~~ iff $ v(x) - v(y) \sim c$
\item  $q \models b$ ~~~~~~~~~~~~~~ iff $b \in L(q)$
\item $q \models \neg \varphi$ ~~~~~~~~~~~~iff $q \not\models \varphi$
\item $q \models \varphi_{1} \lor \varphi_{2}$ ~~~~~ iff $q \models \varphi_{1}$ or $q \models \varphi_{2}$
\item $q \models \varphi_{1} \exists\U \varphi_{2}$ ~~~~ iff for some trajectory $q^{'} \in M$ with $q^{'} (0,0) = q$,

there exists a position $(i, \delta)$ of $q^{'}$ such that $q^{'}(i, \delta) \models_{M, \E+\tau(i,\delta)} \varphi_{2}$ and \\
for all positions $(j, \epsilon)$ of $q^{'}$, if $(j, \epsilon) \prec (i, \delta)$ then $q^{'}(j, \epsilon)\models_{M,\E+\tau(j,\epsilon)} \varphi_{1} \lor \varphi_{2}$.

\item  $q \models \varphi_{1} \forall\U \varphi_{2}$ ~~~~ iff for some trajectory $q^{'} \in M$ with $q^{'} (0,0) = q$,

there exists a position $(i, \delta)$ of $q^{'}$ such that $q^{'}(i, \delta) \models_{M, \E+\tau(i,\delta)} \varphi_{2}$ and \\
for all positions $(j, \epsilon)$ of $q^{'}$, if $(j, \epsilon) \prec (i, \delta)$ then $q^{'}(j, \epsilon)\models_{M,\E+\tau(j,\epsilon)} \varphi_{1} \lor \varphi_{2}$.
\end{itemize}
\qed
\end{definition}

The basic TCTL operator in the above definition is the \textit{until} operator
which can be used to define the time interval in which the property should be true. 
As an example of the use of TCTL, consider the property ``It is always true that
$e_{1}$ may be followed by $e_{2}$ within 10 time units". This property can be expressed
in TCTL as $\AG ~ (e_{1} \rimp [true ~\U_{[0,10]}] ~ e_{2})$. 








\section{UPPAAL Model Checker} \label{sec:four} 
UPPAAL \cite{Beh04} is a model checker for real-time systems developed in conjunction by Uppsala University,
Sweden, and Aalborg University, Denmark. It extends the basic timed
automata with features for concurrency, communication, data variables,
and priority. The current version of the tool is 4.1.4 and it is freely available at
\href{http://www.uppaal.com} {http://www.uppaal.com}.
UPPAAL uses a dense-time model to describe systems, where
each clock variable evaluates to a real number.
An UPPAAL model is a parallel composition of all of its timed automata.
All automata start at its initial state (location) and run
independently of each other unless synchronization with other automata
is required. A transition is enabled when all enabling conditions
are evaluated to true and all the synchronization statements are executed.
If more than one transitions are enabled, one
of them is chosen non-deterministically.
In addition to binary synchronisation, UPPAAL supports
also urgent and broadcast synchronisation.
Synchronisation on urgent channels should occur as soon
as both components are enabled. Note that in transitions
with urgent channels, guards can not have clock constraints.
A broadcast channel allows a process (component) to synchronise
with more than one component at the same time. Note that
broadcast channels in UPPAAL are non-blocking in the sense that if a process
has a transition with $a!$, where $a$ is declared
as a broadcast channel, then $a!$ can be performed even when no
$a?$ action is enabled on any of the processes of the system. 


UPPAAL uses a client-server architecture which splits
the tool into a graphical user interface (client) and a model checking engine (server).
The user interface consists of three main sections: system
editor, simulator, and verifier. The editor allows the user to model
the system as a network of timed automata. The simulator gives
the user the capability to interactively run the system to check if there are some trivial errors in
the system design. 
The verifier allows the user to enter the properties to be verified
in a restricted language of CTL.
UPPAAL can verify safety, bounded liveness, and reachability properties.
UPPAAL uses fragment of TCTL language and it does not support the direct verification of
bounded response properties.

\subsection{The Protocol in UPPAAL}
\begin{figure}[h]
  \begin{minipage}[b]{0.5\linewidth}
    \centering
    \includegraphics[width= 2.4in]{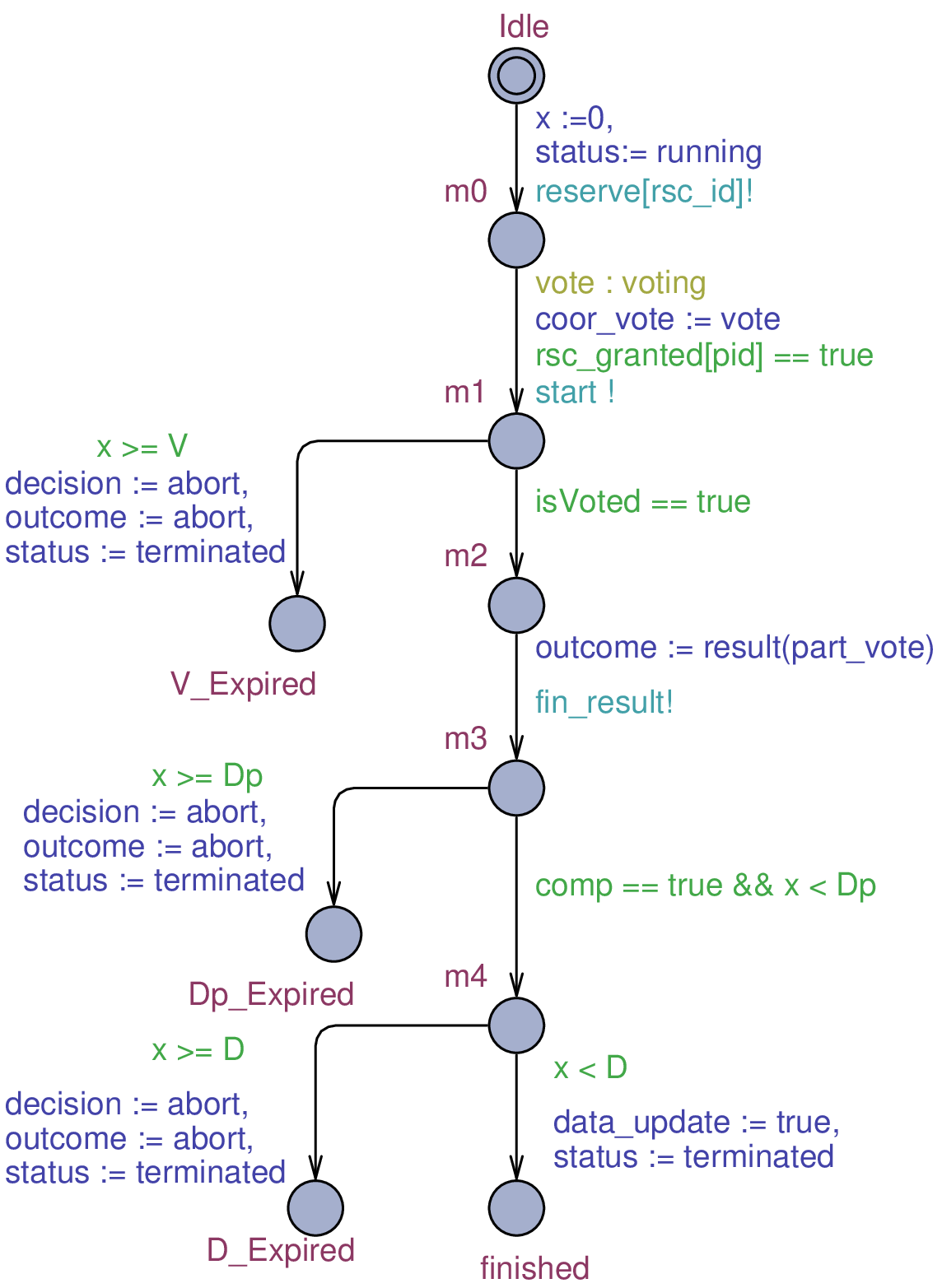}
    \caption{The coordinator template}
    \label{fig:coordinator}
  \end{minipage}
  \hspace{0.5cm}
  \begin{minipage}[b]{0.5\linewidth}
    \centering
    \includegraphics[width= 2.4in]{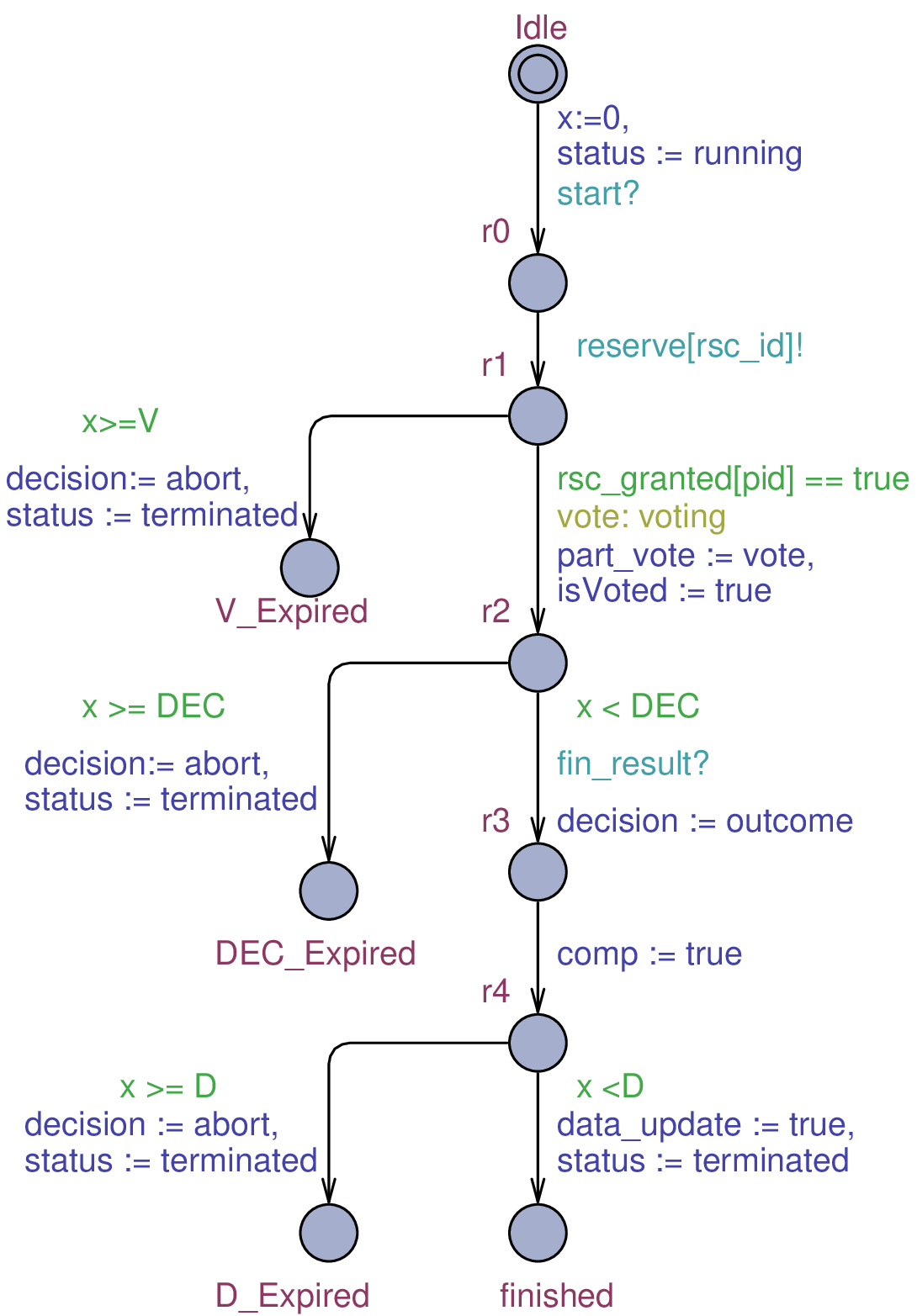}
    \caption{The participant template}
    \label{fig:participant}
  \end{minipage}
\end{figure}

\begin{figure}[h]
  \begin{minipage}[b]{0.6\linewidth}
    \centering
    \includegraphics[width= 2.5in]{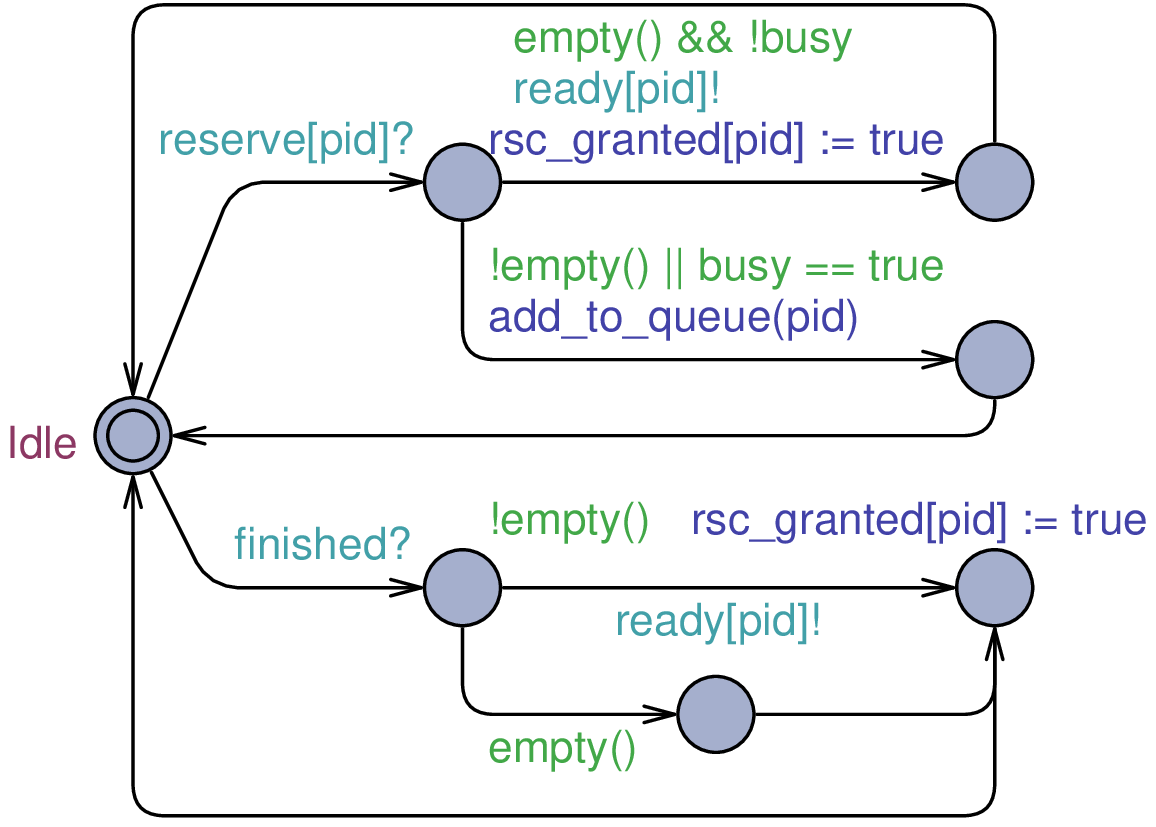}
    \caption{The resource manager template}
    \label{fig:Rscmanager}
  \end{minipage}
  \hspace{0.5cm}
  \begin{minipage}[b]{0.5\linewidth}
    \centering
    \includegraphics[width= 2.5in]{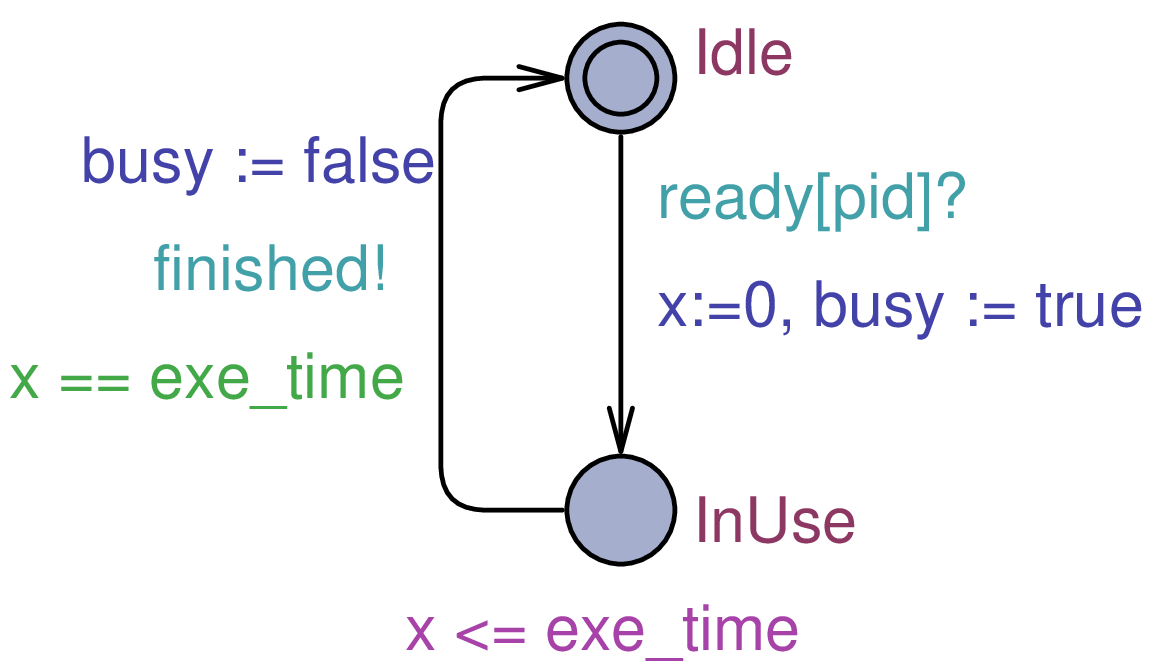}
    \caption{The abstract CPU template}
    \label{fig:resource}
  \end{minipage}
\end{figure}

The coordinator template is depicted in Figure \ref{fig:coordinator}.
Initially, the coordinator attempts to reserve a CPU time slot
via sending a reservation request signal to the CPU resource manager (see Figure \ref{fig:Rscmanager})
using the channel \verb+reserve[rsc_id]+ indexed with the resource to be allocated.
If the CPU is busy in executing other tasks, the manager will add the coordinator process
to the waiting queue. Otherwise, it will send immediately the process to the CPU for processing.
When the manager receives a \verb+finished+ signal from the CPU  
indicates that the CPU has finished processing the current process
and it is currently in an \verb+idle+ state, the manager will send the process at the
front of the queue (if any) to the CPU for processing. 
The abstract model of the CPU (see Figure \ref{fig:resource}) has two locations \verb+idle+
and \verb+InUse+ which reflects the status of the CPU.
When it receives a \verb+ready[pid]+ signal from process \verb+pid+, it moves from 
\verb+idle+ to \verb+InUse+, and then returns from \verb+InUse+
to \verb+idle+ after the determined execution time is completed.
If the resource (CPU) is granted (\verb+rsc_granted ==true+),
the coordinator initiates the protocol via broadcasting a \verb+start+ message to all the participants.
The coordinator then waits to receive the votes of the participants.
If $V$ time units passed before receiving all the votes, 
the coordinator decides to abort and then terminate.
Otherwise, it will move to location $m2$ at which it decides and broadcast the decision.
A function \verb+result(part_vote)+ returns the result of the
transaction based on the values of the received votes. The
coordinator broadcasts this result using the broadcast channel
\verb+fin_result+ and the global variable \verb+outcome+.
The coordinator then moves to location $m3$ at which it waits to receive the completion
messages of the participants. If $D_{p}$ time units passed before receiving
all completion messages, it decides to abort and then terminate.
The protocol ends successfully at location \verb+finished+ at which the coordinator
updates its database server assuming that the clock \verb+x+ 
does not exceed the deadline \verb+D+.

The template of the participants is depicted in Figure \ref{fig:participant}.
All the participants start their execution at location \verb+idle+
where they wait to receive a \verb+start+ signal from the coordinator.
Once they receive that signal, each participant $i$ will try
to reserve $t_{i}$ time units via signalling the resource manager component. 
If the CPU is busy at that time,
it will join the waiting queue until it gets executed. 
If the deadline $V$ expired before sending their votes
to the coordinator they decide to abort and then terminate. 
Each participant then moves to location $r2$ at which 
it waits to receive the decision of the coordinator.
If it does not receive it within $DEC$ time units,
it decides to abort the transaction and terminate. Otherwise, 
it sets its \verb+comp+ variable to true and moves to location $r4$
where it updates its database server and terminates. 

\section{Rabbit Model Checker} \label{sec:five} 
Rabbit \cite{Beyer03} is a model checking tool for real-time systems. 
The theoretical foundation of the tool is mainly based on timed automata 
extended with concepts for modular modeling. This tool has been developed
by Dirk Beyer and his research group at the University of Passau, Germany.
The current version of the tool is 2.1. 
The tool is freely available at 
\href{http://www.sosy-lab.org/~dbeyer/Rabbit/} 
{http://www.sosy-lab.org/~dbeyer/Rabbit}.
The tool deals with timed systems using BDD data structure and uses some heuristic techniques to compute 
good variable orderings of the model and an estimate of the BDD size.
Rabbit has been designed to be able to handle large
 systems in an efficient way and it supports the following features:

\begin{itemize}

\item Modular Modeling. Rabbit uses modular extension of timed automata called Cottbus Timed 
Automata (CTA). That is, the automata are encapsulated by modules.
Each module in the model has an explicit \textit{interface} that contains variables
and synchronisation labels that are used to allow the module to communicate with
other modules. In Rabbit we can declare models in a hierarchical structure in the sense
that modules in the model can contain other modules.

\item Reachability Analysis. Rabbit uses an efficient approach for analysing
reachability properties of timed automata. It uses a symbolic 
approach based on BDD data structure as a representation of the  
explored state space. 

\item Refinement Checking. Rabbit verifies large systems 
via replacing their concrete models with more abstract ones.
It uses simulation-based techniques to show that both models, the 
concrete and the abstract one, have the same behaviour.
\end{itemize}

We now give an informal description of the formalism of Cottbus Timed Automata (CTA),
which is used in the modeling language of Rabbit. \cite{Beyer01}. 

A CTA system consists of a set of modules that can be defined in a hierarchical way.
One of them (the one at the top of the hierarchy)
is used to model the whole system.
 The other modules are used as templates.

Each module in the system model should have the following components:

\begin{itemize}

\item An \textit{identifier}. Identifiers are used to name the modules within the system description.
Using identifiers we can create several instances of the modules associated with these identifiers.

\item An \textit{Interface}. The interface of a module contains the declarations of the variables
that are used in that module. In a CTA module, we can declare clock variables, discrete variables, 
and synchronisation labels.
\begin{itemize}
\item \textit{Synchronisation labels}. Sometimes called signals which are used
to synchronise timed automata that exist in different modules in the system. 
The concept of synchronisation labels in modules is very similar to the concept of events in CSP.
 
\item \textit{Variables}. Rabbit allows us to declare both continuous (clock) variables
and discrete variables. The values of these variables can be 
updated using assignment statements in the transition rules of the automaton.
\end{itemize}

\item A \textit{timed automaton}. Each module contains a timed automaton. 
The automaton consists of a finite set of states, a finite set of transitions,
and a set of synchronisation labels. In CTA language, a process transition is declared 
as a transition rule which starts with the keyword \verb+TRANS+, while the locations
of the automaton are declared using the keyword \verb+STATE+.

\item \textit{Initial condition}. This is a formula over the module variables and the states
of the module's automaton, which specifies the initial state of the module.

\item \textit{Instances}. In CTA model, a module can contain instances of of the other defined 
modules in the model. This is useful when we model systems containing subsystems, and that
these subsystems occur several times in a system.

\end{itemize}
 
 The textual description of the protocol in the Rabbit input
language is given in Appendix \ref{Rabbit}.


\section{RED Model Checker} \label{sec:six} 

RED \cite{wang04} stands for (Region-Encoding Diagrams) is a TCTL model checker for real-time systems.
This model checker has been designed and developed by Farn Wang and colleagues at the National Tawian University.
The current version of the tool is 5.0. The RED tool is available for free at \href{http://cc.ee.ntu.edu.tw/~farn/}{http://cc.ee.ntu.edu.tw/~farn}.
An interesting feature of the RED model checker is that it is totally based on symbolic technology with BDD-like diagrams. 
RED now supports varieties of data-structures, which can be described as follows. 
\begin{itemize}

\item Timed automata with CRD (Clock-Restriction Diagram) technology, 

\item linear hybrid automata with HRD (Hybrid-Restriction Diagram) technology, 

\item MDD-based diagrams which stands for multi-value decision diagrams and is very much like BDD except
 that variables are now decimal and arcs are labeled with value intervals.

\end{itemize}

In RED, systems are described as parametrized communicating timed automata (CTA),
where processes can be model processes, specification processes, or environment
processes. In a system with $n$ processes, the user invokes the RED model checker via telling it which
processes are for the model, and which for the specification. The remaining processes
will be for the environment. Since the automata in RED are parametrised automata 
(i.e. we can pass parameters to them) then we can declare many process automata with 
the same automaton template and identify each process automaton with a process index.
RED supports both forward and backward analyses, deadlock detection, and
counter-example generation. In RED, users can declare global and local
variables of type boolean, discrete, clock-restriction variable, and hybrid-restriction variable.
The tool also supports several optimisation (reduction) techniques that help to
manipulate the state space of the system under consideration in an efficient way such as symmetry reduction, inactive
variable elimination, performance enhancement techniques for time inevitability evaluation.
RED differs from the two tools discussed above in that it uses an extended version of TCTL with.
It also supports strong and weak fairness assumptions of CTA. 

The textual description of the protocol in the language of RED is given in Appendix \ref{RED_model}.

\section{Correctness Conditions}
\label{sec:seven} 

The first formula of interest
is global atomicity  (i.e. all processes must agree on the final decision: all must abort or all must commit.) 
We might first specify atomicity as $ \textbf{AFAG} ~ (\bigwedge_{i\neq j}~ (i.\verb+decision+ = j.\verb+decision+))$.
Intuitively, the formula states that for all possible paths there will be a state where all processes
globally agree on their final decision. 
Unfortunately, the protocol can not be expected to satisfy this since processes might decide
to commit and then change their decision to abort if the deadlines \verb+Dp+ or \verb+D+ 
expired during the execution of the protocol:
suppose that all the participants have voted to commit the transaction and they were able to deliver
their votes before the expiry of the deadline \verb+V+.
Suppose further that the coordinator broadcasted a commit message before
the expiry of the \verb+DEC+ deadline but it failed to receive the completion messages
within the deadline \verb+Dp+. In this case,
the coordinator will change the global decision from $\commit$ to $\abort$
and then broadcast an abort message. So there are situations
in the protocol where processes may change their decisions.
We therefore write the following weaker specification of atomicity: 

\textit{ Specification 1: The global atomicity is always guaranteed.} 
 $$ \AG ~ (\bigwedge_{i\neq j} ~ \neg  (i.\verb+decision+ = \abort \land j.\verb+decision+ = \commit))$$

Note that the variable \verb+decision+ can take one of
the following values $\{ \textit{undefined}$, \abort, \commit $\}$.
Initially, we set the decision of each participant to the \textit{undefined} value.
Note that by verifying atomicity property we verify implicitly 
consistency property since ensuring atomicity guarantees data 
consistency on all distributed database systems.

The second property of interest is validity (i.e. if process $i$ 
votes `no' then `no' is the only possible decision value,
also if all processes vote `yes' then `yes' is the only decision value).
We might specify validity as follows:

\textit{ Specification 2(a): If one process voted `no' then all processes will eventually abort.} 
$$ 
\AG ~(\bigvee_{i=1..n} (i.\verb+vote+ = \no) \rimp \AF ~(\verb+outcome+ = \abort))
$$
\textit{ Specification 2(b): If all processes voted `yes' then all processes will eventually commit.} 
$$
\AG~  (\bigwedge_{i=1..n} (i.\verb+vote+ = \yes)  \rimp \AF ~ (\verb+outcome+ = \commit) )
$$

Recall that the goal of the protocol is to preserve data consistency as well
as to satisfy all designated intermediate 
deadlines $D_{p}$, $DEC$, and $V$. If any of these deadlines expired 
during the execution of the transaction, it is immediately killed 
(i.e. all processes will decide to abort).
Note that the execution of the transaction may be delayed due to 
queuing delay or due to a communication delay 
which might cause the protocol to miss its deadlines. 
The following three specifications verify whether the protocol can satisfy these deadlines.

First, the participants should respond in a bounded time delay to the coordinator
commit request message via sending their votes no more than $V$ time units after the coordinator
sends the request.

\textit{Specification 3: If there are no faults and the coordinator $C$ sent a commit request message
at time $x_{1}$, then it is guaranteed to receive all participants' votes by $x_{1}+V$ on the the coordinator's clock.}
$$
\begin{array}[t]{l}
    \AG ~ ((\verb+C.request_sent+ = \textit{true} ~\land ~\verb+t+ =x_{1}) \rimp \\
   \AF ~(\bigwedge_{i=1..n} ~(\verb+C.vote_received[i]+ = \textit{true}) ~ \land ~ \verb+t+ \leq x_{1} + V))
\end{array}
$$
Second, the coordinator should broadcast the decision no more than
$DEC$ time units after receiving the participants votes.

\textit{Specification 4: If there are no faults and the coordinator received the votes 
at time $x_{2}$, then all the participants can receive the decision by $x_{2}+DEC$ on any participant $P_{i}$'s clock.}
$$
\begin{array}[t]{l}
  \AG ~ ((\bigwedge_{i=1..n} (\verb+C.vote_received[i]+ = \textit{true}) ~\land ~ \verb+t+ =x_{2})) \rimp \\
  \AF ~ (\bigwedge_{i=1..n} (i.\verb+decision_received+ = \textit{true}) ~ \land ~ \verb+t+ \leq x_{2} + DEC))
\end{array}
$$

Next, the coordinator should be able to receive all acknowledgements within $D_{p}$
time units after it sent the decision.

\textit{Specification 5: If there are no faults and the coordinator sent the decision
at time $x_{3}$, then it can receive all acknowledgements within $x_{3}+ D_{p}$ on the coordinator's clock.}
$$
\begin{array}[t]{l}
 \AG ~ ((\verb+C.decision_sent+ = \textit{true} ~\land ~\verb+t+ =x_{3}) \rimp \\
  \AF ~ ( \bigwedge_{i=1..n} ~(\verb+C.ack[i]+ = \textit{true}) ~ \land ~ \verb+t+ \leq x_{3} + D_{p}))
\end{array}
$$
Finally, all processes should terminate within $D$ time units after
the coordinator initiates the protocol. Note that we consider the protocol failed
to achieve its goal if it can not completed within $D$ time units after it started.

\textit{Specification 6: The protocol must always terminate within $D$ time units after it started.}

$$ \AG ~ (\verb+t+ = D  \rimp  \bigwedge_{i=1 ..n} (i.\status = \verb+terminated+))$$


\section{Specifying The Protocol Properties in  UPPAAL, Rabbit, and RED} 
\label{sec:eight}

In this section, we discuss how we verify the properties of the protocol
in the specification language of each tool. UPPAAL uses fragment of TCTL logic,
RED uses complete TCTL logic, while on other hand, TCTL is not available in Rabbit 
and it uses techniques based on reachability analysis to verify systems properties.
In order to simplify the discussion of the specifications, we consider
a setting with only one coordinator and one participant. However,
in our verification of the protocol (see section \ref{sec:comparison}),
we consider also settings with large numbers of participants.

In UPPAAL, we can capture specification 1 as follows.
$$
\verb+A[] not (coor.decision == commit and part.decision == abort)+
$$
The formula states that it is always not (never) the case that one process
will decide to commit and one other process decides to abort (i.e. all processes must
agree on the final decision). 

Since Rabbit does not support the TCTL language, it alternatively provides an analysis command language to write a simple
segment of code for verifying properties based on reachability analysis. 
Using this language, we declare a set of variables that are used to represent
a set of states, called regions, followed by a set of iterative command statements. 
We then check whether the model can reach a region 
where the formula can be violated.
The code below expresses specification 1 in Rabbit's analysis language.
\begin{verbatim}
REACHABILITY CHECK T2PC
{
1   VAR  initial, error, reached : REGION;
2   COMMANDS
3   initial:= INITIALREGION;
4   error := ((coor.decision == 1) AND (part.decision ==2)) OR 
5            ((coor.decision == 2) AND (part.decision ==1));
6   reached := REACH FROM initial FORWARD; 	
7   IF (EMPTY(error INTERSECT reached)){
8     PRINT "Specification 1 satisfied.";}
9   ELSE  { PRINT " Specification 1 violated.";} 
}
\end{verbatim}
The first line declares three regions. 
Region \verb+initial+ represents the set of initial states from the Rabbit's modules (see appendix \ref{Rabbit}).
Lines 4 and 5 characterize the set of states that violate specification 1 of the protocol:
some process decided to abort while some other process decided to commit. 
Line 6 assigns to \verb+reached+ the set of states reachable from the initial state.
The specification is satisfied if the intersection between the \verb+reached+ region
and the \verb+error+ region is empty. Note that we verify the protocol model
using the \verb+FORWARD+ reachability analysis option. One can verify the model also
using backward reachability via replacing the option \verb+FORWARD+ in line 
6 with the \verb+BACKWARD+ option.

In RED we can express specification 1 as follows.
$$
\verb+forall always not (decision[1] == 1 && decision[2] == 2)+
$$
 The above RED formula says that for all possible paths of the protocol, there is 
 no path where process 1 (the coordinator process) decides to abort (\verb+decision[1] == 1+)
 while process 2 (the participant process) decides to commit (\verb+decision[2] == 2+).

Specification 2 can be captured in UPPAAL's specification language as follows.

\begin{verbatim}
A [] (coor_vote == abort or part_vote == abort imply outcome == abort) or 
     (coor_vote == commit and part_vote == commit imply outcome ==  commit)
\end{verbatim}
Intuitively, the above UPPAAL formula says that if one of the processes
voted to abort the transaction then the global outcome of the protocol will be abort. While on the other hand, if
all processes voted to commit then the final outcome will be commit.

In Rabbit we can verify specification 2 as follows.
\begin{verbatim}
REACHABILITY CHECK T2PC
{
1   VAR  initial, error, reached : REGION;
2   COMMANDS
3   initial:= INITIALREGION;
4   error := ( ((part.vote == 1) OR (coor.vote == 1)) AND (outcome == 2)) OR 
5            ( (part.vote == 2) AND (coor.vote ==2) AND (outcome == 1));
6   reached := REACH FROM initial FORWARD; 	
7   IF (EMPTY(error INTERSECT reached))
8      {PRINT "Specification 2 satisfied.";}
9   ELSE { PRINT " Specification 2 violated.";}  }
\end{verbatim}
Lines 4 and 5 in the above Rabbit code characterize the set of states 
that violate specification 2: one process voted to abort the transaction while the coordinator
decided to commit it, or all processes voted to commit the transaction while the 
coordinator decided to abort it. Again the specification is satisfied
if the intersection between the \verb+reached+ region and the \verb+error+ region
is empty.

While in RED we can specify property 2 as follows.
\begin{verbatim}
 forall always ((vote[1] ==1 || vote[2]==1 implies outcome ==1) OR 
                (vote[1] ==2 && vote[2]==2 implies outcome ==2))
\end{verbatim}
Note that the local variable \verb+vote+ can take one of the following values 0 (undefined), 1 (abort), and 2 (commit).

Next, since UPPAAL and Rabbit lack the expressiveness to capture directly bounded response properties,
we therefore verify specifications (3-5) via reducing them into reachability properties. 
We can verify specification 3 in the tools UPPAAL, Rabbit, and RED
respectively as follows.

$$
\verb+A[] not (coor.V_Expired+)
$$
The formula states that there is no path where process
\verb+coor+ can reach location \verb+V_Expired+. 
If the property does not satisfy then there exists a counterexample
where the deadline \verb+V+ expires before the coordinator can receive
the votes of the participants. 

In  Rabbit we express specification 3 as follows.
\begin{verbatim}
REACHABILITY CHECK T2PC
{
1   VAR  initial, error, reached : REGION;
2   COMMANDS
3   initial:= INITIALREGION;
4   error := STATE (Coor.coor) = waitVotes AND coor.x > V;
5   reached := REACH FROM initial FORWARD; 	
6   IF (EMPTY(error INTERSECT reached)){
7     PRINT "Specification 3 satisfied.";}
8   ELSE  { PRINT " Specification 3 violated.";} 
}
\end{verbatim}
The declaration of the region \verb+error+ in the above Rabbit code states that the clock of coordinator
exceeds the deadline \verb+V+ while the coordinator is still waiting for the participant's vote at location 
\verb+waitVotes+.

Since RED uses complete TCTL language, we can then express specification 3 straightforwardly as follows.
\begin{verbatim}
forall always (waitVotes[1] => forall eventually {<= V} sendDecision[1])
\end{verbatim}
Intuitively, the above RED formula says that the coordinator process (process 1 in the RED model) can
always move from location \verb+waitVotes+ to location \verb+sendDecision+ within \verb+V+ time units.
According to the RED model 
(see appendix \ref{RED_model}), this means that the coordinator can always receive the participants'
votes within \verb+V+ time units after it sends the \verb+start+ message.

Next we can express specification 4 in UPPAAL as follows.
$$
\verb+A[] not (part.DEC_Expired+)
$$
We can capture specification 4 in Rabbit language using a code similar to the code that
we use to express specification 3 while modifying the assignment of the \verb+error+ region as follows.
\begin{verbatim}
        error := STATE(Part.part) = waitDEC AND part.x > DEC;
\end{verbatim}

In RED we can express the specification as follows.
\begin{verbatim}
forall always (part_wait[2] => forall eventually {<= DEC} sendCompMsg[2])
\end{verbatim}

Specification 5 can be expressed in UPPAAL as follows.
$$
\verb+A[] not (coor.Dp_Expired+)
$$
To verify specification 5 in Rabbit we modify the assignment of the \verb+error+ 
region as follows.
\begin{verbatim}
       error := STATE(Coor.coor) = waitCompMsg AND coor.x > Dp;
\end{verbatim}
In RED we specify the property as follows.
\begin{verbatim}
forall always (waitCompMsg[1] => forall eventually {<= Dp} coor_final[2])
\end{verbatim}
Finally, specification 6 can be verified in UPPAAL as follows.

\begin{verbatim}
       A<> (coor.x == D imply coor.status == terminated  and 
            part.x == D imply part.status == terminated)            
\end{verbatim}
In Rabbit we verify specification 6 as follows.
\begin{verbatim}
REACHABILITY CHECK T2PC
{
1   VAR  initial, error, reached : REGION;
2   COMMANDS
3   initial:= INITIALREGION;
4   error := ((coor.x >= D) AND (coor.status == 0)) OR 
5            ((part.x >= D) AND (part.status == 0));
6   reached := REACH FROM initial FORWARD; 	
7   IF (EMPTY(error INTERSECT reached))
8      {PRINT "Specification 6 satisfied.";}
9   ELSE { PRINT " Specification 6 violated.";}  
}
\end{verbatim}
Lines 4 and 5 in the above Rabbit code characterize the set
of states that violate specification 6: the coordinator
clock \verb+x+ exceeds the deadline \verb+D+ while the coordinator is still running,
or the participant clock \verb+x+ exceeds the deadline
while the participant process is still running.

In RED we can specify the property as follows.
\begin{verbatim}
      forall eventually (x[1] == D implies status[1] == 1 AND 
                         x[2] == D implies status[2] == 1) 
\end{verbatim}

Note that the local variable \verb+status+ in the above RED's formula can take one of the following 
values 0 (means the process is running), and 1 (means the process has been terminated).

\section{Verification Results} \label{sec: results}
We verify first a strong formula of atomicity $\AG (\bigwedge_{i\neq j} ~ (i.\verb+decision+ = j.\verb+decision+)) $
in which all processes can reach to the same decision at the same time. 
Recall that initially each process sets its decision to the \textit{undefined} value which will
be changed during the execution of the protocol to either \textit{commit} or \textit{abort} based on the values of the votes.
As we expect the formula does not satisfy since processes may not be able to receive the decision
of the coordinator at the same time due to communication delay or queuing delay.
We therefore consider a weaker definition of atomicity $ \textbf{AFAG} ~ (\bigwedge_{i\neq j}~ (i.\verb+decision+ = j.\verb+decision+))$.
We also found this specification fails since as we discussed before processes 
might change their decision from commit to abort if the deadlines \verb+Dp+ or \verb+D+
expire during the execution of the protocol. We finally verify atomicity as expressed in \textit{specification 1}
and found the specification holds. Under this weak definition of atomicity the
consistency of the databases might be temporarily broken. For example we might
have situations in the protocol where process $i$ commits at real-time 10,
process $j$ commits at real-time 15, and process $k$ commits at real-time 20.
The protocol allows such situations to occur, in
particular when some processes get delayed while waiting 
to get executed. However, when the protocol does terminate,
the consistency of the distributed databases is preserved.

The next question of interest is then whether validity specifications always hold , as
claimed. The answer obtained by model checking is that \textit{Specification 2(a)} holds
while surprisingly \textit{Specification 2(b)} fails where the counter-example discovered is the following:

\textbf{Example 1:} Suppose that all processes have voted to commit the transaction 
($ \bigwedge_{i=1..n} (i.\verb+vote+ = \yes)$)
and that the coordinator has received the votes within the deadline \verb+V+. 
Suppose further that the coordinator broadcasted the global decision before
the expiry of the deadline \verb+DEC+ but it
failed to receive the completion messages within the deadline \verb+Dp+.
According to the rules of the protocol, the coordinator will decide
to abort the transaction due to the expiry of \verb+Dp+ and
set the global variable \verb+outcome+ to the $\abort$ value.
Hence we have a situation where although all processes voted to commit,
the coordinator decided to abort the transaction, and therefore
\textit{Specification 2(b)} fails.

The correctness of specifications (3-6) depends mainly upon
the satisfaction of the following three conditions: 
(1) absence of node or link failures since failures if happen might cause some processes to miss their deadlines.
(2) the ability of the operating environment to deliver messages on-time and that
	messages loss is not possible.
(3) fair scheduling policy: each process $i$ is guaranteed to execute at least
     $t_{i}$ time units: the time needed to receive a decision message from the coordinator, perform
     the commit or abort action, and send a completion message to the coordinator.
     In our modeling of the protocol we assume that all resource managers use
     non-preemptive scheduling policy in the sense that when a process starts
     executing, it continues executing until it completes its determined execution time.
Since we made these three assumptions about the environment of the protocol,
we therefore found specifications (3-6) hold.

\section{Model Checking Performance} \label{sec:comparison} 

In this section, we present the model checking runtimes obtained
in testing the tools, with version 4.1.4 for UPPAAL, 2.1 for Rabbit,
and 5.0 for RED. All experiments are conducted on a PC with Redhat Linux 7.3 with Intel (R)
core(TM)2 Quad CPU 2.6 GHz and 512 MB memory.
The specifications of the protocol were checked with backward
and forward analysis in Rabbit and RED, and using the on the fly approach
for UPPAAL. In the forward analysis method the model checker attempts
to construct a characterization of all states that can be reached from a set of initial states $I$
with respect to the declared behaviour structure $M$. In the backward reachability analysis
the model checker attempts to construct a characterization of all states that can reach  
specific goal states (unsafe states) with respect to the declared behaviour structure.
While in on-the-fly approach, the state space of the system is generated dynamically
and only the minimal amount of information is stored in memory.
In the tables below we show the CPU
time used by the system on behalf of the calling process (system time).
An entry of ``x" indicates that the model checker ran out of memory
on that specification.
Rabbit and RED ask for only one input file which describes both
the components of the system and the formula to be verified. 
While on the other hand, UPPAAL asks for two input files,
one for the system and another one for the logical formula.
As shown in section \ref{sec:seven} some properties of the protocol require us to use
a full TCTL language and verify formulas with nested temporal modalities which are not allowed in both UPPAAL and Rabbit. 
We have therefore weakened the properties of the protocol
until UPPAAL and Rabbit can handle them, as shown in section \ref{sec:eight}.

\begin{table}[h]
\setlength{\tabcolsep}{1pt}
\small
\begin{center} 
\begin{tabular}{|c|c||c|c|c|c|c|c|c|} 
\hline 
\multicolumn{9}{|c|}{\textbf{Backward analysis}} \\
\hline
\multicolumn{2}{|c|}{} & \multicolumn{7}{c|}{Specification} \\
\hline
Number of processes & Model Checker & 1 & 2(a) &2(b) & 3 &4 &5&6 \\  
\hline 
  6 & Rabbit &1.22 &1.25 & 1.66& 1.21& 1.37& 1.5& 1.68\\ 
  6 & RED & 10.88 & 4.13 & 8.42 & 12.9 & 11.26 & 9.57 & 6.19\\ 
  \hline
  9 & Rabbit& x& x& x& x& x& x&x\\ 
  9 & RED&554 &249 & 734& 981& 859& 732&1584\\ 
  \hline 
  12 & Rabbit& x&x&x&x&x&x&x\\ 
  12 & RED&2667 & 979& 5524& 6135& 6283& 4339& 8540\\ 
  
\hline 
\end{tabular} 
\end{center} 
\caption{Model Checking Runtimes (seconds) for Rabbit and RED With Backward Analysis \label{table:backward}}
\end{table} 

\begin{table} [h]
\setlength{\tabcolsep}{1pt}
\small
\begin{center} 
\begin{tabular}{|c|c||c|c|c|c|c|c|c|} 
\hline 
\multicolumn{9}{|c|}{\textbf{Forward analysis}} \\
\hline
\multicolumn{2}{|c|}{} & \multicolumn{7}{c|}{Specification} \\
\hline
Number of processes & Model Checker & 1 & 2(a) & 2(b)& 3 &4 &5&6 \\  
\hline 
  6 & Rabbit& 160& 162& 164& 160& 161&163 &172\\ 
  6 & RED & 2.58& 1.69&2.47 & 1.19&1.36 &1.52 &1.87 \\ 
  \hline
  9 & Rabbit& x& x& x& x& x& x&x\\ 
  9 & RED&69.7 &46.3 &59.3 &26.9 &29.5 &31 &40.6\\ 
  \hline 
  12 & Rabbit& x& x& x& x& x& x&x\\ 
  12 & RED& 3088& 1705& 1989& 884& 939& 943&1322\\ 
\hline 
\end{tabular} 
\end{center} 
\caption{Model Checking Runtimes (seconds) for Rabbit and RED With Forward Analysis \label{table:forward}}
\end{table} 

We scaled the model of the protocol until the tools could not 
verify the protocol properties, due to state space problem.
The complexity of the analysis depends mainly on the number of clocks
and states in the model. 
In Table 1 we give the runtimes obtainted in checking the protocol using Rabbit backward reachability analysis
and RED backward TCTL model-checking. 
RED could verify successfully the protocol up to 12 processes with 8 clocks,
while Rabbit could verify only the simplest cases of the protocol. 
Although that Rabbit uses a discrete-time semantics of timed automata which is
considered easier and simpler than the
dense-time semantics adopted by RED and UPPAAL, it failed to analyse instances
of the protocol with large number of processes.
In Table 2 we report the runtimes obtained in testing the tools Rabbit
and RED using forward reachability analysis. Optimizations
used in RED make it more scalable than Rabbit by several order of magnitude.
It should be noted that the latest available version of Rabbit has been released 
in 2003 which has not been updated since that date. While on the other hand,
many optimisations have been added to the tool
RED in the last few years, which give it the 
opportunity to perform much better than Rabbit.

In Table 3 we give the model-checking runtimes of the protocol
using UPPAAL's on the-fly approach. 
UPPAAL could verify successfully the protocol up to 9 processes with 6 clocks.
As we can see, on the whole, the CRD-based tool RED could analysed the protocol
more efficient than the DBM-based tool UPPAAL and the BDD-based tool Rabbit.

\begin{table}[h]
\setlength{\tabcolsep}{1pt}
\small
\begin{center} 
\begin{tabular}{|c||c|c|c|c|c|c|c|} 
\hline 
\multicolumn{8}{|c|}{\textbf{On The Fly Approach}} \\
\hline
\multicolumn{1}{|c|}{} & \multicolumn{7}{c|}{Specification} \\
\hline
Number of processes & 1 & 2(a) & 2(b) &3 &4 &5&6 \\  
\hline 
  6  & 0.01& 0.008& 0.007& 0.001& 0.002& 0.003& 0.045\\ 
\hline
  9 & 4.4& 0.01& 0.06& 0.05& 0.013& 0.084& 0.01\\ 
  \hline 
  12 &x & 0.9& 0.04& 0.03& 0.53& x &0.057\\ 
\hline 
\end{tabular} 
\end{center} 
\caption{Model Checking Runtimes (seconds) for UPPAAL With On the Fly Approach \label{table:fly}}
\end{table} 

In general this case study increased our insight into the state of the art 
in dense-time model checking, and our understanding of the algorithms
and data structures that are used in the implementation of the current dense-time
model checking tools. It shows also the capabilities and the limitations of the 
model checkers UPPAAL, Rabbit, and RED.

\section{Related Work} \label{sec:relatedwork} 

Some work has already been done on the verification
of commitment protocols using formal techniques.  
In particular, the basic 2PC protocol has frequently been
the focus of studies of verification of distributed computing \cite{jeff,atif09,Jan92,Olve08},
but it is just one of several variants discussed in the literature.
One of the interesting variants of the protocol is the T2PC protocol that has complex timing constraints.
In this work we have shown how the protocol can be
analyzed with three various tools: UPPAAL, Rabbit, and RED.
To the best of our knowledge the T2PC protocol has not been model checked before. 

The literature of timed automata theory is a rich literature
since it was introduced by Alur and Dill in 1990. 
Alur and Madhusudan \cite{Alur04} present a full survey of known results for decidability
problems in timed automata theory. Tripakis \cite{Tripakis98} gives
algorithms and techniques to verify timed systems using TCTL logic and Timed Buchi Automata,
which have implemented in KRONOS model checking tool.
One of the disadvantages of KRONOS is that its input language supports a very restricted
data types that allow only the declaration of clock variables. For this
reason we have not included KRONOS in our comparative study.

In the last few years, BDD-like data structures have been used in the verification
of timed systems. The model checkers Rabbit has been developed based
on BDD technology. Emprical results given in \cite{Farn04} and \cite{Beyer03} have shown
that RED and Rabbit outperformed UPPAAL in some particular examples 
such as Fisher mutual exclusion and FDDI Token Ring protocol.

Beyer and Noack \cite{Dirk03} compared the three tools via considering the verification
of a set of simple benchmarks such as the Fischer mutual exclusion protocol, and CSMA/CD. 
Experiments show that the tools RED and Rabbit perform better than UPPAAL when 
considering a large number of processes. However, heuristics used
in Rabbit make the tool more scalable than RED. 

Wang \cite{Farn04} show that CRDs outperform DBMs when verifying specifications
that contain large number of clocks. However, he pointed out that CRDs consume much
space (memory) in handling intermediate data structures, and therefore
require intensive use of garbage collection.
Note that
the emprical results presented in the above discussed works were reported using an old version of
UPPAAL (v3.2.4), which lack many of the optimisations that are used
in the current version of the tool (v4.1.4).

RED is able to verify properties that are not expressible in UPPAAL and
Rabbit and it supports full TCTL language with fairness assumptions. RED also
allows verifying formulas that contain nested temporal modalities (i.e. $\AF \AG(\varphi)$).
On the other hand, UPPAAL's specification language supports fragment of
TCTL and nested temporal formulas are not allowed, while Rabbit language
is restricted to reachability formulas. Unlike UPPAAL, RED and Rabbit provide no
graphical interface or simulation facilities. One advantage of Rabbit over
the other tools is that it supports modular modelling that allows us to represent
the system components in a hierarchy way. This facilitates the modelling of complex 
timed systems. However, UPPAAL has richer expressiveness in modeling
real-time systems than Rabbit and RED. (i.e. The support of data variables in
Rabbit and RED is more limited than it is in UPPAAL.)

\section{Conclusions} \label {sec:conclusion}
We have modeled and verified the real-time version of the two phase commit protocol
in the model checkers UPPAAL, Rabbit, and RED.
RED outperformed UPPAAL and Rabbit in terms of performance and scalability
and allowed us to verify the protocol for large number of processes. 
Our analysis showed that the central property of the protocol, the data consistency property,
might be temporarily broken during the execution of the protocol,
however, when the protocol terminated the consistency is preserved. 
The three model checkers vary in how easy, or difficult, it is to formalise the protocol
in the language of each model checker, and then to verify its correctness properties.
The input language of UPPAAL, Rabbit, and RED are substantially different, and
they are suited for different classes of examples. We intend to pursue
this investigation by considering more complex real-time commit protocols and verifying
more complex real-timed properties.

\bibliographystyle{plain}
\bibliography{references}

\begin{thebibliography}{10}

\bibitem{Alur98}
R.~Alur.
\newblock Timed automata.
\newblock In {\em NATO ASI Summer School on Verification of Digital and Hybrid
  Systems}. 1998.

\bibitem{Alur93}
R.~Alur, C.~Courcoubetis, and D.~Dill.
\newblock Model-checking in dense real-time.
\newblock In {\em Information and Computation}, pages 2--34. 1993.

\bibitem{Alur90}
R.~Alur, C.~Courcoubetis, and D.~L. Dill.
\newblock Model-checking for real-time systems.
\newblock In {\em Proceeding of the 5th Annual Sympoisum on Logic in Computer
  Science}, pages 414--425. IEEE Computer Society Press, 1990.

\bibitem{Alur94}
R.~Alur and D.~Dill.
\newblock A theory of timed automata.
\newblock In {\em TCS}, pages 183--235. 1994.

\bibitem{Alur04}
R.~Alur and P.~Madhusudan.
\newblock Decision problems for timed automata: A survey.
\newblock In {\em International School on Formal Methods for the design of
  Computer, Communication and Software Systems, SFM-RT 2004}, pages 200--236.
  2004.

\bibitem{atif09}
M.~Atif.
\newblock Analysis and verification of two-phase commit and three-phase commit
  protocols.
\newblock In {\em Emerging Technologies ICET'09}, pages 326--331, 2009.

\bibitem{Beh04}
G.~Behrmann, A.~David, and K.G. Larsen.
\newblock {A tutorial on Uppaal}.
\newblock In {\em {Formal Methods for the Design of Real-time Systems (SFM-RT
  2004)}}, pages 200--236. Springer, 2004.

\bibitem{Ber87}
A.~Philip Bernstein, Vassos Hadzilacos, and Nathan Goodman.
\newblock Concurrency control and recovery in database systems.
\newblock Addison-Wesley, 1987.

\bibitem{Beyer03}
D.~Beyer, C.~Lewerentz, and A.~Noack.
\newblock {Rabbit: A tool for BDD-based verification of realtime systems}.
\newblock In {\em Proceedings of the 15th International Conference on Computer
  Aided Verification (CAV 2003)}, pages 122--125, 2003.

\bibitem{Dirk03}
D.~Beyer and A.~Noack.
\newblock Can decision diagrams overcome state space explosion in real-time
  verification?
\newblock In {\em Proceedings of the 23rd IFIP International Conference on
  Formal Techniques for Networked and Distributed Systems (FORTE 2003)}, pages
  193--208, 2003.

\bibitem{Beyer01}
Dirk Beyer and Heinrich Rust.
\newblock Cottbus timed automata: Formal definition and semantics.
\newblock In {\em Proceedings of the 2nd IEEE/IFIP Joint Workshop on Formal
  Specifications of Computer-Based Systems (FSCBS 2001)}, pages 75--87, 2001.

\bibitem{Dav89}
S.~Davidson, I.~Lee, and V.~Wolfe.
\newblock A protocol for times atomic commitment.
\newblock In {\em Proceeding of 9th International Conference On Distributed
  Computing System}, 1989.

\bibitem{Daws96}
C.~Daws, A.~Olivero, S.~Tripakis, and S.~Yovine.
\newblock {The tool KRONOS}.
\newblock In {\em Hybrid Systems III}. 1996.

\bibitem{prompt}
Jayant~R. Haritsa, Krithi Ramamritham, and Ramesh Gupta.
\newblock The prompt real-time commit protocol.
\newblock {\em IEEE Trans. on Parallel and Distributed Systems}, 11:160--181,
  2000.

\bibitem{Jan92}
W.~Janssen and J.~Zwiers.
\newblock Protocol design by layered decomposition.
\newblock In {\em Proc. FTRTFTS 2nd International Symposium}, pages 307--326.
  LNCS, 1992.

\bibitem{CGP99}
E.~M.~Clarke Jr., O.~Grumberg, and D.~A. Peled.
\newblock {\em Model Checking}.
\newblock The MIT Press, 1999.

\bibitem{kaynar03}
D.~Kaynar, N.~Lynch, R.~Segala, and F.~Vaandrager.
\newblock {Timed I/O Automata: a mathematical framework for modelling and
  analyzing real-time systems}.
\newblock In {\em Proceedings 24th IEEE International Real-Time Systems
  Symposium (RTSS03)}, pages 166--177, 2003.

\bibitem{lam}
K.Y. Lam, J.~Cao, C.L.Pang, and S.H. Son.
\newblock Resolving conflicts with committing transactions in distributed
  real-time databases.
\newblock In {\em {Proceedings of 3rd IEEE International Conference on
  Engineering of Complex Computer Systems (ICECCS'97)}}, 1997.

\bibitem{Larsen97}
K.~Larsen, F.~Larsson, P.~Pettersson, and W.~Yi.
\newblock Efficient verification of real-time systems: Compact data structures
  and state-space reduction.
\newblock In {\em Proceedings of the 18th IEEE Real-Time Systems Symposium},
  pages 14--24, 1997.

\bibitem{jeff}
Jeff Magee.
\newblock Analyzing synchronous distributed algorithms.
\newblock 2003.

\bibitem{Olve08}
\"{O}lveczky Peter~Csaba.
\newblock {Formal Modeling and Analysis of a Distributed Database Protocol in
  Maude}.
\newblock In {\em Proceedings of the 2008 11th IEEE International Conference on
  Computational Science and Engineering - Workshops}, pages 37--44, 2008.

\bibitem{Qin}
B.~Qin and Y.~Liu.
\newblock High performance distributed real time commit protocol.
\newblock {\em Journal of Systems and Software, Elsevier Science Inc.}, pages
  1--8, 2008.

\bibitem{Tripakis98}
S.~Tripakis.
\newblock {\em The analysis of timed systems in practice}.
\newblock PhD thesis, Universite Joseph Fourier, Grenoble, France, 1998.

\bibitem{Farn04}
F.~Wang.
\newblock {Efficient verification of timed automata with BDD-like data
  structures}.
\newblock {\em International Journal on Software Tools for Technology Transfer
  (STTT)}, pages 77--97, 2004.

\bibitem{wang04}
F.~Wang.
\newblock Model-checking distributed real-time systems with states, events, and
  multiple fairness assumptions.
\newblock In {\em {Proceedings of the 10th International Conference on
  Algebraic Methodology and Software Technology, AMAST 2004}}, pages 553--568,
  2004.

\end{thebibliography}

\appendix
\section {The Textual Description of The Protocol in Rabbit} \label{Rabbit}
 \begin{verbatim}
MODULE Coordinator
{
   LOCAL
      x : CLOCK; vote : DISCRETE(3) ; update : DISCRETE;
      decision : DISCRETE(3) ; status : DISCRETE;
   INPUT
       D : CONST;    //the deadline of the transaction.
       V : CONST;   // the deadline for receiving participants' votes.
       Dp: CONST;  //  the deadline for receiving participants' completion messages.
       commit :   SYNC;  // to commit the transaction.
       abort  :   SYNC; // to abort the transaction.
       comp   :   SYNC; // to receive the participant's completion message
  OUTPUT
      reserve   : SYNC;   // to reserve a CPU time slot.
      start     : SYNC;  //  to start the T2PC protocol.
  MULTREST
     outcome : DISCRETE; 
     resource_granted : DISCRETE; 
     dec_sent : DISCRETE;
  // Set initial state of automaton and clock values.
INITIAL
   STATE(Coor) = init AND x = 0 AND status =0;
AUTOMATON Coor
{
   STATE init{ TRANS{ SYNC ! reserve;   GOTO begin;}
                         TRANS {GUARD x >= V;     GOTO fail;} }
   STATE begin{ TRANS{ GUARD resource_granted = 1; 
                       SYNC ! start; DO vote' =1; GOTO waitVotes;}
                TRANS  {GUARD resource_granted = 1; 
                        SYNC ! start; DO vote' =2; GOTO waitVotes;}}
   STATE waitVotes  {TRANS{  SYNC ?abort; DO decision' =1 AND dec_sent' =1
                             AND outcome' =1;  GOTO waitCompMsg;}
                     TRANS  {GUARD vote =2; SYNC ?commit; DO decision' =2 AND
                             dec_sent' =1 AND outcome' =2; 
                             GOTO waitCompMsg;}
                     TRANS  {GUARD vote =1; SYNC ?commit; DO decision' =1 AND
                             dec_sent' =1 AND outcome' =1; 
                             GOTO waitCompMsg;}
                     TRANS  {GUARD x >= V; GOTO fail;}}
  STATE waitCompMsg{  TRANS {SYNC ?comp; GOTO finish; }
                      TRANS {GUARD x >= Dp; GOTO fail;}}
 
  STATE finish{  TRANS  {GUARD x <=D; DO update' = 1 AND status' =1;
                         GOTO exception;} 
                 TRANS  {GUARD x> D; GOTO exception;} } 
  STATE fail{    TRANS  {DO decision' = 1 AND outcome' =1 
                         AND status' =1; GOTO exception;}} 
  STATE exception {}
}
}
// The following module represents the template of the participant.
MODULE Participant 
{
 LOCAL
     x : CLOCK; vote : DISCRETE (3); 
     decision : DISCRETE(3) ; status : DISCRETE(2); update : DISCRETE;  
 INPUT 
     D : CONST;  V : CONST ;  DEC : CONST;  start : SYNC; 
 OUTPUT      
      reserve :  SYNC;  commit :  SYNC;   abort :  SYNC;  comp :  SYNC; 
 MULTREST
     outcome : DISCRETE; 
     resource_granted : DISCRETE; 
     dec_sent : DISCRETE;
 // Set initial state of automaton and clock values.
 INITIAL
     STATE(Part) = init AND x = 0 AND status =0;
AUTOMATON Part
{
  STATE init{   TRANS {SYNC? start;  GOTO reserveTimeSlot;}
                TRANS  {GUARD x >= V; GOTO fail;} }
  STATE reserveTimeSlot{  TRANS  {SYNC ! reserve; GOTO sendVote;}
                          TRANS  {GUARD x >= V;   GOTO fail;} }
  STATE sendVote{  TRANS {GUARD resource_granted = 1 AND x <V;
                          SYNC !abort; DO vote' = 1;  GOTO waitDec;}
                   TRANS {GUARD resource_granted = 1 AND x <V; 
                          SYNC !commit; DO vote' = 2; GOTO waitDec;}
                   TRANS  {GUARD x >= V; GOTO fail;}}  

  STATE waitDec{   TRANS  {GUARD x >= DEC; GOTO fail;}
                   TRANS  {GUARD x < DEC AND dec_sent=1; DO decision' = outcome; 
                           GOTO sendCompMsg;} } 
  STATE sendCompMsg{  TRANS  {SYNC ! comp; GOTO finish;} }
  STATE fail{         TRANS  {DO decision' = 1 AND status' =1;  GOTO exception;} }
  STATE finish{       TRANS  {GUARD x <D; DO update' = 1 AND status' =1;
                              GOTO exception;}
                      TRANS  {GUARD x>= D; GOTO exception;} } 
  STATE exception {}
}
}
// The following module represents an abstract model of the CPU.
MODULE resource 
{
LOCAL
  x : CLOCK;
INPUT
  ready : SYNC;
  exe_time : CONST;
OUTPUT
  finished : SYNC;
MULTREST
  busy : DISCRETE; // 0 means the CPU is free, and 1 means it is busy.
INITIAL
   STATE (CPU) = Idle AND x =0 AND busy =0;
AUTOMATON CPU 
{
  STATE Idle{  TRANS  {SYNC ? ready; DO x' =0 AND
                       busy' =1; GOTO InUse;} } 
  STATE InUse{  INV  x<= exe_time; 
               TRANS  {GUARD x = exe_time; SYNC ! finished;  
               DO busy' = 0; GOTO Idle;} } 
}
}
// The template of the CPU manager.
MODULE ResouceManager
{
LOCAL 
   wait : DISCRETE; 
INPUT 
   reserve : SYNC; finished : SYNC; 
OUTPUT
  ready : SYNC;
MULTREST 
  busy : DISCRETE; // 0 means the CPU is free, and 1 means it is busy.
  resource_granted : DISCRETE; 
INITIAL
   STATE (Manager) = Idle;
AUTOMATON Manager
{
  STATE Idle { TRANS  {SYNC ? reserve; GOTO M1;}
               TRANS  {SYNC ? finished; GOTO M2;} }
  STATE M1{  TRANS {GUARD busy =0;
                    SYNC !ready; DO resource_granted' = 1; GOTO Idle;}
             TRANS {GUARD busy = 1; 
                    DO wait' =1; GOTO Idle;} }  
  STATE M2{  TRANS  {GUARD wait =1;  SYNC !ready; DO wait' =0
                     AND resource_granted' = 1; GOTO Idle;}
             TRANS {GUARD wait = 0;  GOTO Idle;}}
}
} 
\end{verbatim}
We pick some statements in the Rabbit model in order to explain 
how to declare the model behaviour structure with Rabbit.
The declaration is a sequence of \verb+STATE+ declarations.
A typical \verb+STATE+ declaration can be found from the statements (S1) to (S3).
The statements declare a state whose name is \verb+InUse+
and whose invariance condition is ``\verb+x<= exe_time+''. 
Inside the transition \verb+TRANS+ 
we have a synchroniser \verb+finished+, a triggering condition ``\verb+x == exe_time+",
and two actions ``\verb+DO busy' = 0;"+ and ``\verb+GOTO Idle;+".
Another statement that we believe it needs to be explained is 
the declaration statement \verb+decision: DISCRETE(3)+. The parameter in the statement
tells Rabbit how many values we want to store
in the variable \verb+decision+. 
The statement indicates that the \verb+decision+ variable can take one of the following values 
$\{0,1, 2\}$.
This gives Rabbit the opportunity to spend
the correct number of bits in the BDD representation.
Note that synchronization between automata is done via synchronization labels and the semantic is as in CSP,
i.e., one automaton can take a transition with label S, if all other automata also take a transition with S.

\section{ The Textual Description of The Protocol in RED}   \label{RED_model}

 \begin{verbatim}
#define D 80
#define D_p 52 
#define DEC 40 
#define V 15 
#define exe_time 52
process count = 6; 
/* One local clock. */
local clock x;
/* A set of synchronizers. */
global synchronizer start, reserve1, reserve2, yes, no, commit, abort,
                    comp, ready1, ready2, finished1, finished2; 
local discrete vote: 1..2;
local discrete decision: 0..2;
local discrete update: 0..1;
local discrete status: 0..1; 
global discrete busy1: 0..1;
global discrete busy2: 0..1;
global discrete outcome: 1..2;
global discrete resource_granted1: 0..1;
global discrete resource_granted2: 0..1;
global discrete wait1: 0..1; 
global discrete wait2: 0..1; 
/* 7 modes for the coordinator. */
 mode coor_idle (true) 
 { 
   when ! reserve1 (true) may goto coor_begin;
 }
  mode coor_begin (true) 
 { 
   when !start (resource_granted1 ==1)  may x= 0; vote =1; goto wait_votes;
   when !start (resource_granted1 ==1)  may x= 0; vote =2; goto wait_votes;
 }
 mode wait_votes (x<=D) 
 { 
   when ?yes (vote ==2 && x < V)  may decision =2; outcome =2; goto sendDecision; 
   when ?yes (vote ==1 && x < V)  may decision =1; outcome =1; goto sendDecision; 
   when ?no  (x < V) may decision = 1; outcome =1; goto sendDecision;  
   when (x >= V) may goto coor_fail;
 }
 mode sendDecision (x <=D)
 {
   when !commit (decision == 2) may goto waitCompMsg;
   when !abort  (decision == 1) may goto waitCompMsg;
 } 
 mode waitCompMsg (x<=D) 
 {
   when ?comp (x < D_p) may goto coor_final; 
   when (x>D_p) may goto coor_fail; 
 }  
 mode coor_final (true) 
 {
   when (x < D) may update =1; status =1;
   when (x> D) may goto coor_fail;
 }
 mode coor_fail (true) 
 { 
   when (true) may decision =1; outcome =1; status =1;
 } 
/* The behaviour of the participant can be described as follows.*/
 mode part_idle (true) 
 { 
   when ?start (true) may x = 0; goto part_reserve; 
 }

 mode part_reserve(true)
 {
    when ! reserve2 (true) may goto part_start;
 }
 mode part_start (x < D) 
 { 
   when !yes (resource_granted2 ==1 && x < V) may vote =2; goto part_wait;
   when !no  (resource_granted2 ==1 && x < V) may  vote =1; goto part_wait;
   when (x >= V) may goto part_fail;
 } 
 mode part_wait (true) 
 {
   when ?abort (x < DEC) may decision =1; goto sendCompMsg;
   when ?commit (x < DEC) may decision =2; goto sendCompMsg;
   when (x>= DEC) may goto part_fail;
 }
 mode sendCompMsg (true)
 {
   when !comp(true) may goto part_final;
 }
  mode part_final (true) 
 {
   when (x < D) may update =1; status =1;
   when (x> D) may goto part_fail;
 }
 mode part_fail (true) 
 { 
   when (true) may decision =1; status =1;
 } 

/* The template of the CPU1 */
 mode CPU1_idle (true)
 { 
   when ? ready1 (true) may busy1 = 1; x =0; goto CPU1_InUse;
 }
 mode CPU1_InUse (x<= exe_time)
 { 
   when !finished1 (x == exe_time) may busy1 = 0; goto CPU1_idle;
 }
 mode CPU2_idle (true)
 { 
   when ? ready2 (true) may busy2 = 1; x =0; goto CPU2_InUse;
 }
 mode CPU2_InUse (x<= exe_time)
 { 
   when !finished2 (x == exe_time) may busy2 = 0; goto CPU2_idle;
 }
 /* The template of the CPU1 manager */
 mode Manager1_idle (true)
 {
   when ?reserve1 (true) may goto M11;
   when ?finished1 (true) may goto M21;
 }
 mode M11 (true)
 {
   when !ready1 (busy1 ==0) may resource_granted1 =1; goto Manager1_idle;
   when (busy1 ==1) may wait1 =1; goto Manager1_idle;
 }
 mode M21 (true)
 {
   when !ready1 (wait1 ==1) may wait1 =0; resource_granted1 =1; goto Manager1_idle;
   when  (wait1 ==0)  may goto Manager1_idle;
 } 
 /* The template of the second CPU2 manager */
 mode Manager2_idle (true)
 {
   when ?reserve2 (true) may goto M12;
   when ?finished2 (true) may goto M22;
 }
 mode M12 (true)
 {
   when !ready2 (busy2 ==0) may resource_granted2 =1; goto Manager2_idle;
   when (busy2 ==2) may wait2 =1; goto Manager2_idle;
 }
 mode M22 (true)
 {
   when !ready2 (wait2 ==1) may wait2 =0; resource_granted2 =1; goto Manager2_idle;
   when  (wait2 ==0)  may goto Manager2_idle;
 } 
\end{verbatim}

We also pick some statements in the RED model in order to explain 
how to declare the model behaviour structure with RED. The declaration
is a sequence of \verb+mode+ declarations. A typical \verb+mode+
declaration can be found from the statements (S1) to (S2).
The statements declare a mode
whose name is \verb+InUse+ and whose invariance condition ``\verb+x<= exe_time+".
Inside the transition rule \verb+when+ we have 
a synchroniser \verb+finish+, a triggering condition ``\verb+x == exe_time+",
and the two actions ``\verb+busy = 0+" and ``\verb+goto idle+".
The initial condition of the protocol declares six processes. The first process represents
the coordinator database server which starts at \verb+coor_idle+ location,
the second process represents the participant server which starts at the \verb+part_idle+ location,
the third and the fourth processes represent the CPU at each server, and finally
processes 5 and 6 represent the resource manager on each database server.
\end{document}